\documentclass[fleqn,usenatbib]{mnras}

\usepackage[T1]{fontenc}

\DeclareRobustCommand{\VAN}[3]{#2}
\let\VANthebibliography\thebibliography
\def\thebibliography{\DeclareRobustCommand{\VAN}[3]{##3}\VANthebibliography}

\usepackage{graphicx}
\usepackage{amsmath}
\usepackage{amssymb}
\usepackage{cuted}
\usepackage{enumerate}
\usepackage{enumitem}
\usepackage{multirow}
\usepackage{booktabs}
\usepackage{makecell}
\usepackage{tabularx}
\usepackage{bbold}
\usepackage{bm}
\usepackage{xcolor}

\definecolor{valecol}{rgb}{0,0.5, 1.}

\setlist[enumerate]{wide=0pt, widest=99,leftmargin=\parindent, labelsep=* }
\newcommand{\rf}[1]{Fig.~\ref{fig:#1}}
\newcommand{\rt}[1]{Table~\ref{tab:#1}}

\newcommand{\rsssec}[1]{\S\ref{subsubsec:#1}}
\newcommand{\ec}[1]{Eq.~(\ref{eq:#1})}
\newcommand{\eql}[1]{\label{eq:#1}}
\newcommand\be{\begin{equation}}
\newcommand\ee{\end{equation}}
\newcommand\bea{\be\begin{aligned}}
\newcommand\eea{\end{aligned}\ee}

\newcommand{\sfig}[2]{
	\includegraphics[width=#2]{#1}}
\newcommand{\Sfig}[2]{
	\begin{figure}
		\sfig{#1.pdf}{\columnwidth}
		\caption{#2}
		\label{fig:#1}
	\end{figure}}
\newcommand{\Swide}[3]{
	\begin{figure*}
		\sfig{#1.pdf}{#2}
		\caption{#3}
		\label{fig:#1}
	\end{figure*}}

\title[Fast comparison of covariance matrices]{A fast and reliable method for the comparison of covariance matrices}

\author[Ferreira and Marra]{
Tassia Ferreira$^{1,2}$
and Valerio Marra$^{2,3,4,5}$
\\
$^{1}$PPGCosmo, Universidade Federal do Espírito Santo, 29075-910, Vitória, ES, Brazil\\
$^{2}$Laboratório Interinstitucional de e-Astronomia - LIneA, 20921-400, Rio de Janeiro, RJ, Brazil\\
$^{3}$Núcleo de Astrofísica e Cosmologia \& Departamento de Física, Universidade Federal do Espírito Santo, 29075-910, Vitória, ES, Brazil\\
$^{4}$INAF -- Osservatorio Astronomico di Trieste, via Tiepolo 11, 34131, Trieste, Italy\\
$^{5}$IFPU -- Institute for Fundamental Physics of the Universe, via Beirut 2, 34151, Trieste, Italy}

\date{Accepted 2022 May 03. Received 2022 April 01; in original form 2021 December 15}

\pubyear{2021}

\begin{document}
\label{firstpage}
\pagerange{\pageref{firstpage}--\pageref{lastpage}}
\maketitle

\begin{abstract}
Covariance matrices are important tools for obtaining reliable parameter constraints. Advancements in cosmological surveys lead to larger data vectors and, consequently, increasingly complex covariance matrices, whose number of elements grows as the square of the size of the data vector. The most straightforward way of comparing these matrices, in terms of their ability to produce parameter constraints, involves a full cosmological analysis, which can be very computationally expensive. Using the concept and construction of compression schemes, which have become increasingly popular, we propose a fast and reliable way of comparing covariance matrices.
The basic idea is to focus only on the portion of the covariance matrix that is relevant for the parameter constraints and quantify, via a fast Monte Carlo simulation, the difference of a second candidate matrix from the baseline one. To test this method, we apply it to two covariance matrices that were used to analyse the cosmic shear measurements for the Dark Energy Survey Year 1. We found that the uncertainties on the parameters change by 2.6\%, a figure in  agreement with the full cosmological analysis.
While our approximate method cannot replace a full analysis, it may be useful during the development and validation of codes that estimate covariance matrices. Our method takes roughly 100 times less CPUh than a full cosmological analysis.
\end{abstract}




\section{Introduction}
\label{sec:intro}

Cosmology has entered its golden era, with the fast advancement of technology allowing us to build telescopes capable of exploring almost the entire observable universe.
As we brace ourselves for the unprecedented amount of data that will be made available in upcoming surveys like the Javalambre Physics of the Accelerating Universe Astrophysical Survey\footnote{\href{http://www.j-pas.org}{www.j-pas.org}} (J-PAS), the Vera C. Rubin Observatory Legacy Survey of Space and Time\footnote{\href{https://www.lsst.org}{www.lsst.org}} (LSST), Euclid\footnote{\href{https://www.euclid-ec.org}{www.euclid-ec.org}} and the Square Kilometre Array\footnote{\href{https://www.skatelescope.org}{www.skatelescope.org}} (SKA) we must tackle the issue of how to process and extract as much information as possible from the data.

This brings us to the issue of code development and validation~\citep[see, e.g.,][]{Chisari2018,Blanchard2019}. Here, we discuss covariance matrix validation, which is becoming increasingly important in cosmological analyses \citep{Friedrich2021, Krause2017, Joachimi2020}.
Covariance matrices are vital pieces to the puzzle as they take into consideration not only the statistical and systematic errors of the measurement, but also the correlation between each quantity. The size of a covariance matrix grows as the square of the size $N$ of the dataset, which makes them progressively harder to obtain, whether analytically or through simulations. Further, analysing and comparing them also becomes exceedingly difficult.

The most certain and forward way of comparing covariance matrices is in terms of their ability to reproduce cosmological constraints, that is, by employing a full Bayesian analysis. Performing this analysis for every new covariance matrix, however, is very time consuming and computationally expensive. \citet{Friedrich2021}, using the Gaussian linear model,  studied in detail the impact of different covariance matrices on the uncertainties in parameter estimation, the position of the best-fit parameters, and the relative $\chi^2$ value.
In this work, we seek a fast way of comparing covariance matrices that eliminates the need for a cosmological analysis to identify the differences between their constraints.
In other words, we focus on the impact on the uncertainties in parameter estimation.
The motivation is simple: if two covariance matrices produce similar results, then they should, at some level, have comparable features. We find here that even matrices with elements differing by several orders of magnitude (both diagonal and off-diagonal terms) produce parameter estimates that are almost indistinguishable from each other. It is, therefore, conceivable to think that a fair comparison cannot rely on the elements of the full covariance matrix.
In~\cite{Ferreira2020} one of the highlighted results is the potential in using compressed covariance matrices for comparison. It is shown that two different compressed covariance matrices with consistent parameter constraints also show good agreement in a one-to-one element comparison.
Compression methods are powerful tools for reducing the dimensionality of the covariance matrix in order to facilitate and potentially speed up the process of parameter estimation. The most successful compression schemes are capable of taking a covariance matrix of size $N \times N$ and shrinking it down to $n \times n$, where $n$ is the number of free parameters. In this work, we use the Massively Optimised Parameter Estimation and Data compression algorithm (MOPED), as described by~\cite{Heavens1999}, which works remarkably well in the case of Gaussian data, where the model for the mean only depends linearly on the parameters. When these circumstances are met, the method is said to be lossless in the sense that there is no loss of precision in the parameter constraints.

On the other hand, compression schemes are non-invertible, which means that, given a compressed matrix, we are unable to return to the original covariance matrix. In this sense, even if we were able to obtain a compressed covariance matrix, either analytically or with simulations, we would have no way of applying it to our real dataset of interest.

This work has two main goals: i) to develop a method to recreate the full matrix given a compressed one, and ii) to propose a fast and reliable method to compare compressed covariance matrices that discards the need for a full cosmological analysis.

This paper is organised as follows. We start by introducing our data vector and covariance matrices in Section~\ref{sec:cosmic_shear_example}. Then, in Section~\ref{sec:compression_scheme}, we review the matrix compression method we adopt, while we discuss in Section~\ref{sec:rotation} invertible compression. We present our method to compare covariance matrices in Section~\ref{sec:covariance}, where we also analyse two cosmic shear covariance matrices. We conclude in Section~\ref{sec:conclusion}.


\section{Cosmic shear}
\label{sec:cosmic_shear_example}

Light from distant galaxies is deflected by the gravitational field of large-scale structures as it travels through the Universe. This creates a correlated distortion of images, known as cosmic shear, which can be used to directly probe the underlying dark matter distribution and provide insight into the growth of structures and the geometry of the Universe. Cosmic shear has thus emerged as a powerful probe for dark energy~\citep{Kilbinger2015, Hikage2019, Asgari2020, Secco2022}. We review cosmic shear statistics in Appendix~\ref{ap:shear}.

\subsection{The DESY1 data}
\label{subsubsec:DESdata}

The cosmic shear measurements for the Year 1 release of the Dark Energy Survey,~\citep[DESY1,][]{Troxel2017}, were taken over an area of 1321 deg$^2$ of the southern sky and are divided into four tomographic redshift bins from $0.20 < z < 1.30$~\citep{Zuntz2018}, according to the posterior of the photometric redshift as estimated from $griz$ flux measurements~\citep{Boyle2018}.

Each of the 10 bin-pair combinations contains 20 angular bins between 2.5 and 250 arcmin, yielding a data vector of length 200, for each statistic. Not all angular bins are used however, due to cuts that remove angular scales sensitive to baryonic effects, thus reducing the data vector to 167 points for $\xi_{+}(\theta)$and 60 for $\xi_{-}(\theta)$, totalling 227 points. 

We assume a flat $\Lambda$CDM model, with six free parameters, $\left\{A_s, \Omega_m, \Omega_b, \Omega_{\nu}h^2, H_0, n_s\right\}$, and fix $w = -1$ and $\tau = 0.08$. Since we disconsider the baryonic effects, the astrophysical systematics are largely dominated by intrinsic alignment (IA), which describes the coherent orientation of galaxies due to interactions with the underlying gravitational tidal field regions. We vary the amplitude of the nonlinear alignment model, $A_{\text{IA}0}$, and its redshift evolution, $\eta_{\text{IA}}$, which are related via $A_{\text{IA}0} \equiv A_{\text{IA}0}\left[(1 +z)/(1 +z_0)\right]^{\eta_{\text{IA}}}$, where $z_0 = 0.62$ is the pivot redshift. We also have the shear multiplicative bias, $m^i$ which varies with each tomographic bin. Lastly, we vary the photo-$z$ bias, $\Delta z^i$, on the distribution of galaxies in each redshift bin. The priors for these 16 parameters are given in~\rt{priorsDES}. For brevity, we show the posterior probability density functions (PDFs) only for the matter density parameter $\Omega_m$ and the amplitude of matter fluctuations $S_8 \equiv \sigma_8 (\Omega_m/0.3)^{0.5}$.

Finally, the parameter constraints are obtained with \texttt{CosmoSIS}~\citep{Zuntz2015}, while employing the \texttt{MultiNest}~\citep{Feroz2009} sampler and following the same pipeline described in~\citet{Troxel2017}, with the modified likelihood for the transformed and compressed datasets and covariance matrices used in~\citet{Ferreira2020}. The \texttt{MultiNest} run had 1000 \texttt{livepoints}, \texttt{efficiency} set to 0.05, \texttt{tolerance} to 0.1 and \texttt{constant efficiency} set to True.

\begin{table}
	\caption{List of the priors used in the analysis for parameter constraints using the dataset described in~\rsssec{DESdata}. $\mathcal{U}$ denotes flat in the given range and $\mathcal{G}$ is Gaussian with mean equal to its first argument and dispersion equal to its second.}
	\begin{tabular} { l c}
	\hline
		Parameter & Prior	\\ \hline
		Cosmological & \\ [1ex]
		$\Omega_m$ & $\mathcal{U}(0.1, 0.9)$ \\
		$\log A_s$ & $\mathcal{U}(3.0, 3.1)$ \\
		$H_0 \ (\text{km s}^{-1}\text{Mpc}^{-1})$ & $\mathcal{U}(55, 91)$\\
		$\Omega_b$ & $\mathcal{U}(0.03, 0.07)$ \\
		$\Omega_\nu h^2$ & $\mathcal{U}(0.0005, 0.01)$ \\
		$n_s$ & $\mathcal{U}(0.87, 1.07)$ \\ [1ex]
		\hline
		Astrophysical & \\ [1ex]
		$A_{\text{IA}0}$	& $\mathcal{U}(-5, 5)$ \\
		$\eta_{\text{IA}}$& $\mathcal{U}(-5, 5)$ \\ [1ex]
		\hline
		Systematic      & \\ [1ex]
		$m^i$			& $\mathcal{G}(0.012, 0.023)$ \\
		$\Delta z^1$	& $\mathcal{G}(-0.001, 0.016)$ \\
		$\Delta z^2$	& $\mathcal{G}(-0.019, 0.013)$ \\
		$\Delta z^3$	& $\mathcal{G}(0.009, 0.011)$ \\
		$\Delta z^4$	& $\mathcal{G}(-0.018, 0.022)$ \\ [1ex]
        \hline
	\end{tabular}
	\label{tab:priorsDES}
\end{table}

\subsection{The KiDS-1000 data}
\label{subsubsec:KIDSdata}

The measurements for the Kilo-Degree Survey 1000~\citep[KiDS-1000,][]{Asgari2020} contain 1006 deg$^2$ of images, with the primary images taken in the $r$-band, but with the final set having photometry in $ugriZYJHK_s$~\citep{Wright2019}, after being combined with infrared data from the VISTA Kilo-degree INfrared Galaxy survey (VIKING,~\cite{Edge2013}). The data is divided into five tomographic bins, $z_{\text{B}}$, based on their best-fitting photometric redshifts and ranging from $0.1 < z < 1.2$.

There are nine angular bins between 0.5 and 500 arcmin, resulting in a data vector of length 270. The angular cuts are applied to $\xi_-(\theta)$, removing scales with $\theta < 4$ arcmin~\citep{Hildebrandt2017}, which leaves 135 data points for $\xi_+(\theta)$ and 90 for $\xi_-(\theta)$. The final dataset has length 235, with a $235 \times 235$ covariance matrix.

Similarly to the DESY1 analysis, we take a flat $\Lambda$CDM model, with $w = - 1$, and five free cosmological parameters, $\left\{S_8, \Omega_c h^2, \Omega_b h^2, h, n_s\right\}$. There are two astrophysical nuisance parameters: the baryon feedback parameter, $A_{\text{bary}}$ and $A_{\text{IA}}$, where, for analyses with this dataset, the latter does not carry a redshift dependence. The mean of the five redshift distributions is also allowed to vary and are correlated through their covariance matrix. Finally, for the analyses with $\xi_+(\theta)$, we have $\delta_c = \pm \sqrt{c_1^2 + c_2^2}$ to account for the uncertainty of the additive ellipticity bias terms, $c_1$ and $c_2$, assuming that they are constants.~\rt{priorsKIDS} shows the parameters varied as well as their priors.

For parameter constraints, we use the \texttt{MultiNest} sampler within \texttt{CosmoSIS}, with the same settings described in the previous section, but with the KiDS Cosmology Analysis Pipeline, KCAP~\citep{Asgari2020}. We use a modified likelihood to account for the transformed and compressed covariance matrices and data vectors.

\begin{table}
	\caption{List of the priors used in the analysis for parameter constraints using the dataset described in~\rsssec{KIDSdata}.}
	\begin{tabular} { l c}
	\hline
		Parameter      & Prior	\\ \hline
		Cosmological   & \\ [1ex]
		$S_8$          & $\mathcal{U}(0.1, 1.3)$ \\
		$\Omega_c h^2$ & $\mathcal{U}(0.051, 0.255)$ \\
		$\Omega_b h^2$ & $\mathcal{U}(0.019, 0.026)$\\
		$h$            & $\mathcal{U}(0.64, 0.82)$ \\
		$n_s$          & $\mathcal{U}(0.84, 1.1)$ \\ [1ex]
		\hline
		Astrophysical     & \\ [1ex]
		$A_{\text{IA}}$	  & $\mathcal{U}(-6, 6)$ \\
		$A_{\text{bary}}$ & $\mathcal{U}(2.0, 3.13)$ \\ [1ex]
		\hline
		Systematic   & \\ [1ex]
		$\delta_z^1$ & $\mathcal{G}(0, 1.0)$ \\
        $\delta_z^2$ & $\mathcal{G}(-0.181, 1.0)$ \\
        $\delta_z^3$ & $\mathcal{G}(-1.110, 1.0)$ \\
        $\delta_z^4$ & $\mathcal{G}(-1.395, 1.0)$ \\
        $\delta_z^5$ & $\mathcal{G}(1.265, 1.0)$ \\
        $\delta_c$	 & $\mathcal{G}(0, 2.3 \times 10^{-4})$ \\ [1ex]
        \hline
	\end{tabular}
	\label{tab:priorsKIDS}
\end{table}

\subsection{The covariance matrices}
\label{subsubsec:covariances}

There are three distinct covariance matrices used in this work, which will be described in this section. 
What we refer to as the \textit{DES Covariance Matrix} (DCM) is the DESY1 cosmic shear covariance matrix obtained with \texttt{CosmoLike}~\citep{Krause2016}. The covariance is largely dominated by the shape-noise and Gaussian components of the covariance, with a halo model framework being used to include the non-Gaussian parts.

We label the second one as the \textit{Gaussian Covariance Matrix} (GCM) was generated with the code used for the KiDS-450 survey~\citep{Kohlinger2017}, but with the same configuration as DCM, and constituting of only the Gaussian contributions to shape-noise, sample variance and the mixed noise-sample variance term. We do not include the non-Gaussian components and the super-sample covariance so that the differences between the covariance matrices and their parameter constraints are accentuated. It is used here along with the DCM to test our algorithm for comparing covariance matrices. It has been shown in~\cite{Ferreira2020} that their parameter constraints are accurately reproduced when compressed with MOPED.

The third and last one is the \textit{KiDS Covariance Matrix} (KCM), which is used for testing our invertible transformation as well as for some of the tests with modifications to one of the blocks of the transformed matrix. It is the same covariance matrix used in the KiDS-1000 survey analysis with cosmic shear.

The second and third covariance matrices are calculated with the code used for the KiDS-450 and KiDS-1000 surveys~\citep{Kohlinger2017,Joachimi2020}.
Both were obtained analytically and follow the procedure in~\cite{Joachimi2008} for obtaining second-order cosmic shear measurements under the assumption that density field is Gaussian, the galaxies are uniformly distributed, and the survey has a straightforward geometry. The main differences between them are their size, with GCM being 227 $\times$ 227 and KCM 235 $\times$ 235; GCM was produced for DESY1 cosmic shear statistics and KCM is the covariance matrix used in the KiDS-1000 survey; finally, GCM is a Gaussian covariance matrix, while KCM contains both Gaussian and non-Gaussian terms.

DCM and GCM are used in Sections~\ref{sec:rotation} and \ref{sec:covariance}, whereas KCM is only used in Section~\ref{sec:rotation}.


\section{Compression Scheme}
\label{sec:compression_scheme}

Consider a dataset represented by the vector $\mathbf{x} = (x_1, x_2, ..., x_N)$ with probability distribution $L(\mathbf{x}; \boldsymbol{\Theta})$, where $\boldsymbol{\Theta} = (\theta_1, \theta_2, ..., \theta_n)$ is a vector of the model parameters. Take the logarithm of the probability distribution, $\mathcal{L} = \text{ln}L$; the first derivative with respect to the model parameters, $\mathcal{L}_{,i}$, known as the score function, tells how sensitive the model is with respect to the parameters. Its second derivative, the Hessian matrix,
\be\eql{hessian}
    \mathbf{H}_{ij} = \mathcal{L}_{,ij}\,,
\ee
describes the correlation of the estimated values of $\theta_i$ and $\theta_j$. The expectation value of the negative of the Hessian, $\mathbf{F}_{ij} = - \big \langle \mathbf{H}_{ij} \big \rangle$, gives us the Fisher information matrix, which is an essential quantity in Bayesian statistics and, as we will see, for compressing covariance matrices. To obtain a compression scheme capable of retaining the highest amount of information, we seek to maximise $\mathbf{F}_{ij}$; to do so, we start with the log-likelihood function for a Gaussian probability distribution,\footnote{We have dropped the additive constant $n$ln$(2\pi)$.}
\be\eql{likelihood}
    2 \mathcal{L} = \text{ln\ det} \mathbf{C + (x - \boldsymbol{\mu})}^t \mathbf{C}^{-1} \mathbf{(x - \boldsymbol{\mu})}\,,
\ee
with both the covariance matrix, $\mathbf{C = \big \langle (x - \boldsymbol{\mu}) (x - \boldsymbol{\mu})}^t \big \rangle$, and $\boldsymbol{\mu} = \big \langle \mathbf{x} \big \rangle$ dependent of the model parameters $\boldsymbol{\Theta}$. We also define the data matrix as
\be
    \mathbf{D \equiv (x - \boldsymbol{\mu}) (x - \boldsymbol{\mu})}^t\,.
\ee
Taking this definition and using $\text{ln\ det} \mathbf{C} = \text{Tr} \left[ \text{ln} \mathbf{C} \right]$, we can express \ec{likelihood} in a simpler form,
\be
    2 \mathcal{L} = \text{Tr} \Big[\text{ln} \mathbf{C + C}^{-1}\mathbf{D} \Big]\,.
\ee
Derivating for $\theta_i$ and $\theta_j$, we find the quantity we wish to maximise:
\be\eql{fisher1}
    \big \langle \mathcal{L}_{,ij} \big \rangle = \frac{1}{2} \text{Tr} \Big[\mathbf{C}^{-1} \mathbf{C}_{,i} \mathbf{C}^{-1} \mathbf{C}_{,j} + \mathbf{C}^{-1} \mathbf{M}_{ij} \Big]\,,
\ee
where $\mathbf{M}_{ij} = \boldsymbol{\mu}_{,i}\boldsymbol{\mu}_{,j}^t + \boldsymbol{\mu}_{,j}\boldsymbol{\mu}_{,i}^t$. 

There are three ways one can proceed from here, with respect to the dependence on the model parameters: 1) there is the more general case, where both the mean and the covariance depend on the model parameters; 2) only the covariance has a dependence, in which case the second term vanishes; and 3) only the mean is dependent, and the first term vanishes. The general case has been tackled by~\cite{Alsing2017}, where they use the score function to derive $n$ compressed statistics of the data. The second case often reduces to a Karhunen-Loéve eigenvalue problem that results in keeping only the linear combinations of the data with the most informative eigenvalues~\citep{Alonso2017}. Finally, the third case is the basis for MOPED~\citep{Tegmark1996, Heavens1999}, which uses linear compression to radically reduce the dataset. Here, we will only be working with the latter, with an extension for other compression methods being planned for future works. In this case, \ec{fisher1} reduces to
\be\eql{fisher}
    \mathbf{F}_{ij} = \big \langle \mathcal{L}_{,ij} \big \rangle = \frac{1}{2} \text{Tr} \Big[\mathbf{C}^{-1} \mathbf{M}_{ij} \Big]\,.
\ee

For the simplest case of only one parameter, we have
\be
    \mathbf{F}_{11} = \boldsymbol{\mu}_{,1}^t \mathbf{C}^{-1} \boldsymbol{\mu}_{,1}\,.
\ee
If we apply a compression of the type $y = \mathbf{b}^t \mathbf{x}$, we can follow these same steps to obtain the Fisher matrix,
\be\eql{CompFisher}
    \mathbf{F}_{11}^{\text{comp}} = \frac{\mathbf{b}^t \mathbf{M}_{11} \mathbf{b}}{\mathbf{b}^t \textbf{C} \mathbf{b}}\,.
\ee
To find an extremum such that $\mathbf{b}^t \textbf{C} \mathbf{b} = 1$, we apply a Lagrange multiplier,
\be
    \boldsymbol{\mu}_{,1}(\mathbf{b}^t \boldsymbol{\mu}_{,1}) = \lambda \textbf{C} \mathbf{b}\,.
\ee
The solution is the eigenvector which, when normalised, gives
\be\eql{CompScheme}
    \mathbf{b}_1 = \frac{\textbf{C}^{-1} \boldsymbol{\mu}_{,1}}{\sqrt{\boldsymbol{\mu}_{,1}^t \mathbf{C}^{-1} \boldsymbol{\mu}_{,1}}}\,.
\ee
Plugging this back in \ec{CompFisher}, we find
\be
    \mathbf{F}_{11}^{\text{comp}} = \boldsymbol{\mu}_{,1}^t \mathbf{C}^{-1} \boldsymbol{\mu}_{,1} = \mathbf{F}_{11}\,,
\ee
which shows that the Fisher matrix is unchanged.

We can further expand \ec{CompScheme} to multiple parameters. By employing a Gram-Schmidt decomposition, we obtain $y_i$'s that are orthonormal, uncorrelated and carry as much information as possible about the parameter $\theta_i$. We then have,
\be
    \mathbf{b}_n = \frac{\textbf{C}^{-1} \boldsymbol{\mu}_{,n} - \sum^{n-1}_{q=1}\left( \boldsymbol{\mu}_{,n}^t \mathbf{b}_q \right)\mathbf{b}_q}
    {\sqrt{\boldsymbol{\mu}_{,n}^t \mathbf{C}^{-1} \boldsymbol{\mu}_{,n}} - \sum^{n-1}_{q=1}\left( \boldsymbol{\mu}_{,n}^t \mathbf{b}_q \right)^2}\,,
\ee
for $q < n$. With this in hand, the resulting weighing matrix \textbf{b} contains $n$ rows, and the covariance matrix is shrunk to $n \times n$.

Note that the compression vector $b$ only depends on the model parameters. If the model is linear in its parameters, then this dependence disappears, and the method becomes lossless.

To use MOPED as described here, we have assumed a Gaussian likelihood and that the fiducial value at which the derivatives are taken are those at the maximum likelihood point. If one does not have prior knowledge of this value, then one could iterate to find it but, as~\cite{Heavens1999} have found, this is often unnecessary. We also trust that the cosmological model we have chosen is the correct one; deviations from a baseline cosmological model can be accounted for by introducing additional weighing vectors, as described in~\cite{Heavens2020}.

In~\cite{Ferreira2020}, it was established that the MOPED compressed DESY1 covariance matrix and dataset could reproduce the original parameter constraints; here we find that the same is true for KiDS-1000. While these results are not new, verification for KiDS-1000 is an important step of the analysis since, as the authors point out, not all compression schemes are capable of reproducing the original parameter constraints. In particular, those that relied on eigenvalues, or the signal-to-noise ratio showed a loss of constraining power on cosmological parameters since the modes relevant to the IA parameters were discarded by these methods.


\section{Invertible Compression}
\label{sec:rotation}

The next step in our analysis is to obtain an invertible transformation, based on the MOPED compression scheme, that is able to reproduce the same parameter constraints we previously obtained with both the original and the compressed covariance matrices. This step is vital for using the compressed covariance matrix, produced analytically or through simulations, with the observed data vector. 

We note here that this procedure will not reduce the size of the covariance matrix, but rather the number of relevant elements. The transformed matrix is then composed of the compressed covariance matrix and some additional terms. By construction, we do not expect the additional elements to alter the parameter constraints, and we show that this is indeed the case. It is less intuitive, however, that, through the inverse of the invertible transformation, it is possible to generate a new covariance matrix, with elements differing by large orders of magnitude, that retain the same constraining power when using the same data vector. Here we show how this can be achieved and we highlight the perils of considering a comparison between the elements of the full covariance matrices.

\Swide{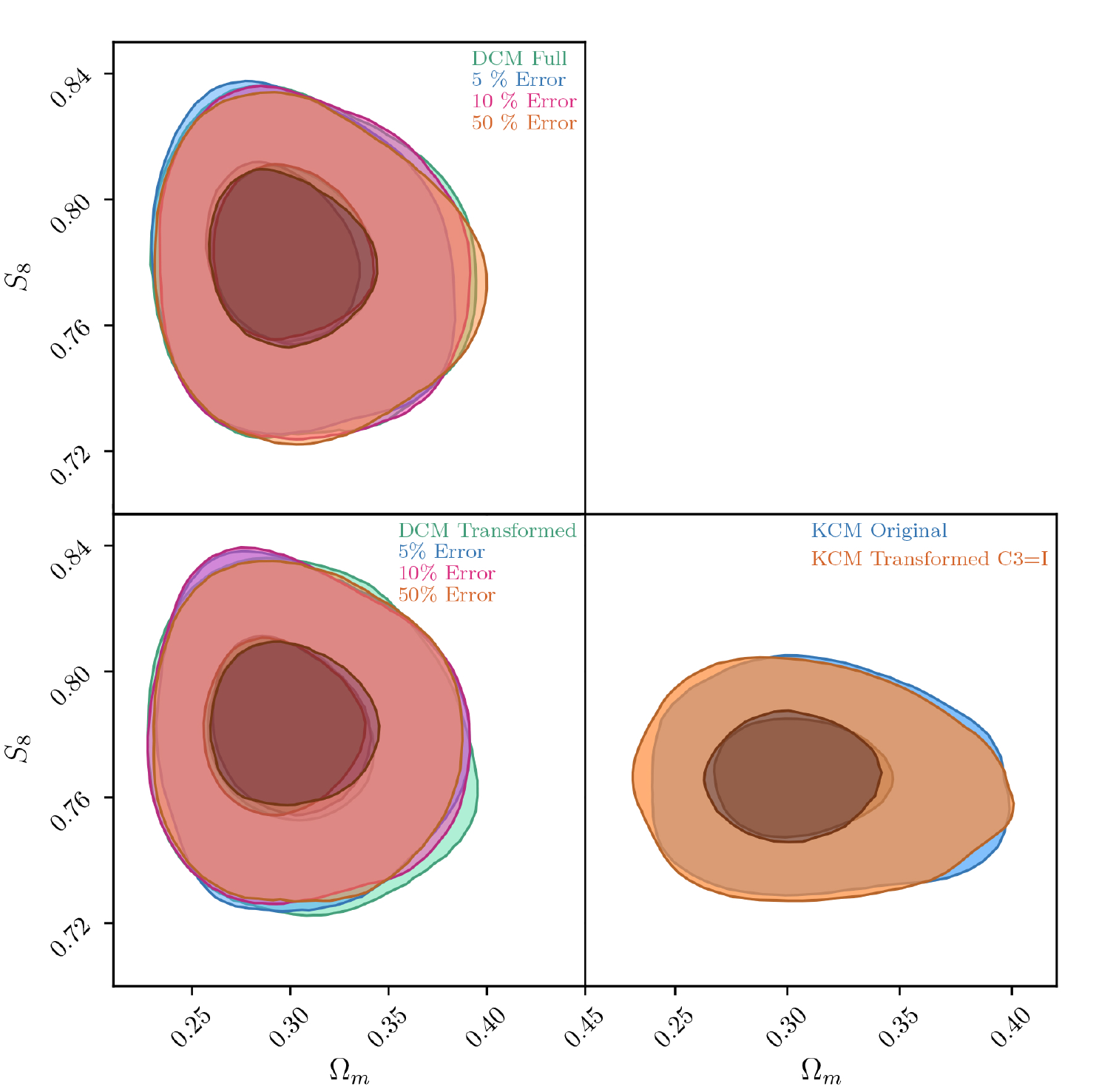}{0.8\textwidth}{\textit{Top:} DESY1 constraints for the original covariance matrix (green) and for three covariance matrices produced by applying the inverse transformation to those with $5\%$ (blue), $10\%$ (pink) and $50\%$ (orange) perturbation applied to the $C3$ block of the transformed DCM.
\textit{Bottom left:} DESY1 constraints on $\Omega_m$ and $S_8$ for the transformed covariance matrix (green) and for a $5\%$ (blue), $10\%$ (pink) and $50\%$ (orange) perturbation applied to the $C3$ block of DCM.
\textit{Bottom right:} KiDS-1000 parameter constraints for the original and transformed covariance matrix for the cosmological parameters $\Omega_m$ and $S_8$. The darker curve, in blue, is for the original covariance matrix and the lighter curve, in orange, is for when the block $C3$ of KCM is replaced by the identity matrix.}

We start with the non-normalised \ec{CompScheme}, and expand it to an invertible, $N \times N$ transformation matrix,
\be
    \mathbf{B} = \left(\mathbf{b} \quad U \right)\,,
\ee
where $U$ has dimension $(N - n) \times N$. We want to find $U$ such that
\be
\mathbf{C}^{\text{trans}} = \mathbf{B}^t \mathbf{C B}
= \left(\begin{matrix}
\mathbf{b}^t \mathbf{C} \mathbf{b} & 0 \\[1.1ex]
0 & U \mathbf{C} U^t
\end{matrix}\right)\,,
\ee
which implies
\be
    \mathbf{b}^t \mathbf{C} U = 0\,.
\ee
For the above to be true, the rows of $U$ must be composed of vectors which form the nullspace of $\mathbf{b}^t \mathbf{C}$. To simplify notation, we represent the transformed covariance matrix partitioned blockwise as
\be\eql{C_trans}
        \left(\begin{matrix}
        C1 & 0 \\[1.1ex]
        0 & C3
\end{matrix}\right)\,.
\ee

In the following Section, we apply this transformation to a toy example, so we can explore, in a more didactic manner, how this transformation alters the covariance matrix.

\subsection{Toy example}
\label{subsec:toy}

In order to better understand our invertible transformation and how $C3$ affects $\mathbf C$, we take a simple toy example. Consider the Gaussian data described by:
\begin{align}
    &\{ t_i, x_i \} \text{ with } i=1,2,3 \,, \nonumber \\
    &C= \sigma^2 I_3 \,, 
\end{align}
where $I_3$ is the 3-d identity matrix and $t_i$ is the independent variable associated to $x_i$. Next, we propose the following model:
\begin{align}
    \mu(t) = \theta_1 + \theta_2 t \,,
\end{align}
for which we can obtain the likelihood as
\begin{align}
    2 \mathcal{L} &= (x_i - \mu(t_i)) C^{-1}_{ij} (x_j - \mu(t_j)) \nonumber \\
    & = \sum_{i=1,2,3} \frac{(x_i - \mu(t_i))^2}{\sigma^2} \,.
\end{align}
The next step is then to derive an explicit expression for \textbf{b}, $U$, $C1$ and $C3$ as a function of $t_i$, $x_i$, $\sigma$ and $\theta_i$. We start with
\be
    \mathbf{b} = \left(\begin{matrix}
    \frac{1}{\sigma^2} & \frac{1}{\sigma^2} & \frac{1}{\sigma^2} \\[1.1ex]
    \frac{t_1}{\sigma^2} & \frac{t_2}{\sigma^2} & \frac{t_3}{\sigma^2}
    \end{matrix}\right)
\ee
and
\be
    U = \left(\begin{matrix}
    - \frac{-t_2 + t_3}{t_1 - t_2} & - \frac{t_1 - t_3}{t_1 - t_2} & 1
    \end{matrix}\right)\,,
\ee
which we can use to find,
\be
    \mathbf{C}^{\text{trans}} = \left(\begin{matrix}
    \frac{3}{\sigma^2} & \frac{t_1+t_2+t_3}{\sigma^2} & 0 \\[1.1ex]
    \frac{t_1+t_2+t_3}{\sigma^2} & \frac{t^2_1+t^2_2+t^2_3}{\sigma^2} & 0\\[1.1ex]
    0 & 0 & \mathbf{C3} \\[1.1ex]
    \end{matrix}\right)\,,
\ee
where
\be
    \mathbf{C3} = 2\sigma^2\ \frac{t_1^2 + t_2^2 + t_3^2 - t_2t_3 - t_1 \left(t_2 + t_3 \right)}{\left(t_1 - t_2 \right)^2}\,.
\ee
Plugging our values in \ec{fisher}, we get,
\be
    F^{\text{trans}}_{ij} = \left(\begin{matrix}
    \frac{3}{\sigma^2} & \frac{t_1+t_2+t_3}{\sigma^2} \\[1.1ex]
    \frac{t_1+t_2+t_3}{\sigma^2} & \frac{t^2_1+t^2_2+t^2_3}{\sigma^2}\\[1.1ex]
    \end{matrix}\right) = C1\,.
\ee
Making $\mathbf{C3} = C3$, we can revert back the transformation to find $C'$. If we substitute $C'$ in \ec{fisher}, we find that $F' = F$, which shows that the Fisher matrix of the modified covariance matrix does not depend on $C3$.

\subsection{Perturbing C3}
\label{subsubsec:C3}

\Swide{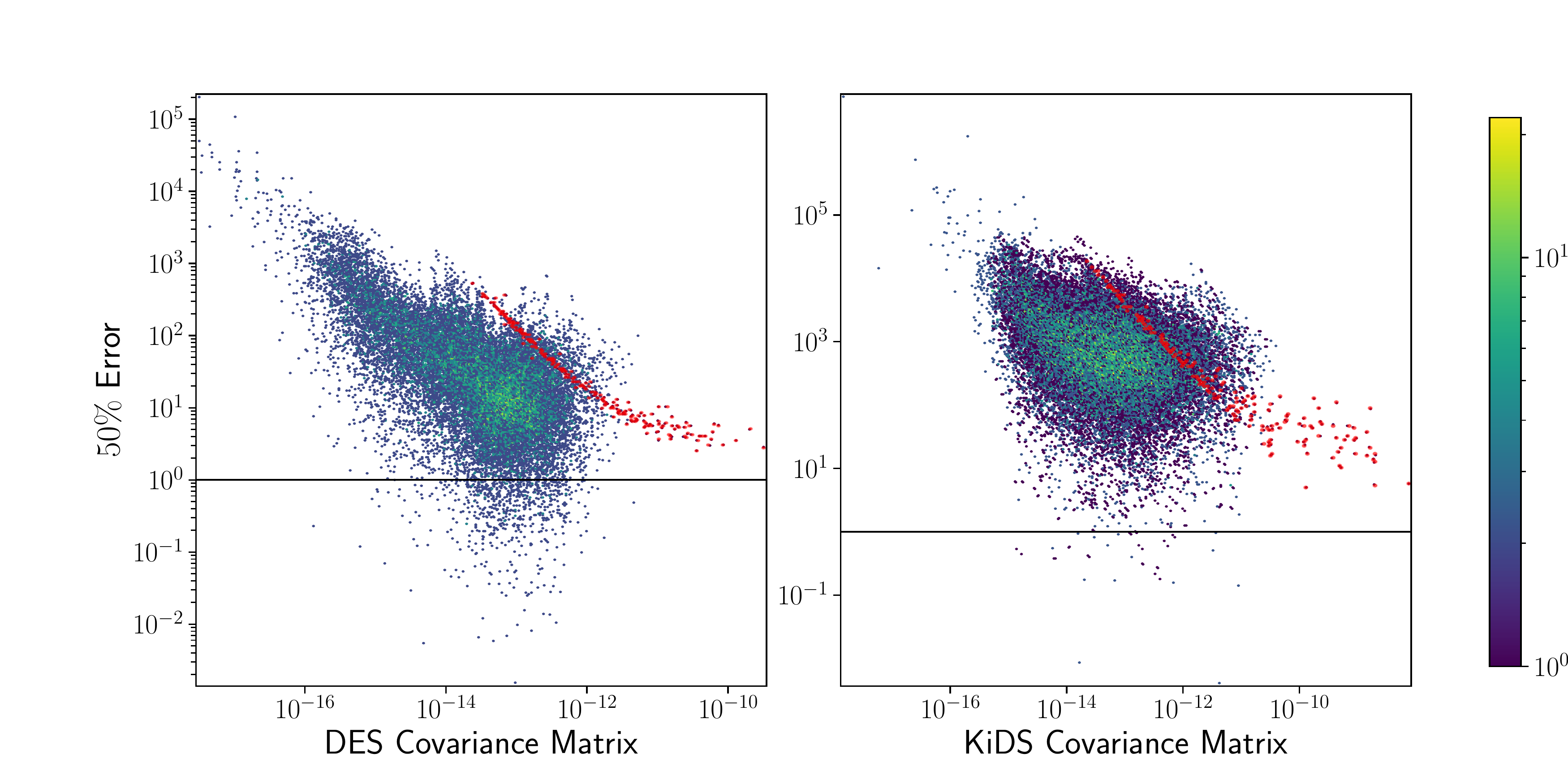}{\textwidth}{Scatter plots of the ratio between the elements of the perturbed covariance matrix with a 50\% Gaussian error and those of the original one. The left panel shows the ratio between the DESY1 covariances, while the right panel shows the ratio relative to KiDS-1000, and the x-axis representing the elements of the respective covariance matrix. The red dots represent the ratio between the diagonal elements of the respective matrices.}

We now describe the ways in which we perturb the $C3$. The impact these modifications have on the parameter constraints are quantified by performing a full cosmological analysis with the new covariance matrix.

We start with the simple task of making $C3 = \mathbb{1}$. This modification increases the diagonal elements by several orders of magnitude and negates all the cross terms of $C3$. We also modify the elements of the $C3$ block of DCM by introducing a Gaussian error of $5\%, 10\%$ and $50\%$ while maintaining its symmetry. In \rf{wmS8-3errors}, we see that these configurations are irrelevant to the parameter constraints, as the contour plots show agreement with the original results.

The next question to address is how this property propagates when we apply the inverse of the transformation \textbf{B} to the perturbed $\mathbf{C}^{\text{trans}}$. We carry out the same procedure for perturbing the elements of $C3$ by $5\%, 10\%$ and $50\%$. We then apply $\mathbf{B}^{-1}$ to transform it into the space of the original data vector. As expected, the results are similar to those found in the previous analysis, see~\rf{wmS8-3errors}.

In~\rf{Ratios} (left for DCM and right for KCM) we show the ratio between the elements of the perturbed covariance matrices and the original one for a $5\%$ (top) and $50\%$ perturbation (bottom), and highlight the diagonal elements in red. A trend can be seen where the smallest elements show a greater disagreement, and this decreases as their values increase. Even the largest elements, however, show a difference of about an order of magnitude. By construction, the broad disparities between the elements should not impact the parameter constraints, and we see that this is indeed the case. It may be, however, that covariance matrices obtained by independent algorithms and show disagreements between their elements, can still produce similar constraints. It is therefore essential to note that metrics to compare covariance matrices based solely on these elements themselves may not produce conclusive results. As such, for an unbiased comparison, we propose to concentrate on the covariance matrices compressed with the MOPED algorithm. 


\section{Comparing Covariance Matrices}
\label{sec:covariance}

We have shown in the previous section that it is imperative that a comparison between covariance matrices be done on its compressed set. This is fortunate because the reduced size results in a speed gain for comparison. In this section we explain the framework we have developed for obtaining a reliable metric of comparison that discards the need for a full cosmological analysis.

Our method separates the comparison into two parts: the analysis of the diagonal elements, the $n$--dimensional vector $\mathcal{D}$, and of the independent elements of the correlation matrix, the $n(n-1)/2$--dimensional vector $\mathcal{C}$. The corresponding calculated differences will then be related to the parameter constraints in terms of their contour levels.
To find the differences we use a Monte Carlo approach by employing Powell's approach for minimising a function~\citep{Powell1964}.

Take two compressed covariance matrices: $C_{\text{base}}$ and $C_{\text{test}}$.%
\footnote{Note that these are the Fisher matrices in the parameters.}
For each step $i$, we create a mock sample $\{\mathcal{D}_{\delta,i}\}$ (or $\{\mathcal{C}_{\delta,i}\}$) by perturbing $\mathcal{D}_{\text{base}}$ (or $\mathcal{C}_{\text{base}}$) with a given error percentage $\delta$.
For the diagonal part, the mocks are generated by drawing $\mathcal{E}_{\mathcal{D}}$ from a multivariate Gaussian distribution $\mathcal{G}[0_n,\delta^2 I_n]$, such that,
\be \label{diago}
    \mathcal{D}_{\delta} = (1+\mathcal{E}_{\mathcal{D}})^2 \, \mathcal{D}_{\text{base}}\,.
\ee
In the case of the correlation matrix, we encounter the restriction that the values must be in the range $[-1, 1]$. Applying this by force could result in a perturbation not cohesive with our chosen $\delta$. This is resolved by switching to the hyperbolic tangent function and correcting for the Jacobian,
\bea
    z &= \tanh^{-1}\mathcal{C}_{\text{base}} \,, \\
    \delta z &= \mathcal{E}_{\mathcal{C}} \cosh^2\left(z + \frac{\mathcal{E}_{\mathcal{C}}}{2} \right) \,,
\eea
where $\mathcal{E}_{\mathcal{C}}$ is drawn similarly to $\mathcal{E}_{\mathcal{D}}$.
Our perturbed vector then becomes,
\be \label{corre}
    \mathcal{C}_{\delta} = \tanh \left(z + \delta z \right)\,.
\ee
Since the hyperbolic function increases asymptotically towards one, we are able to ensure a smooth perturbation.

Next, we produce the sample covariance matrix $S_\delta$ from the generated mocks
\be
    S_\delta = \frac{1}{m-1}\sum_{i=1}^m \left(\mathcal{D}_{\delta,i}-\overline{\mathcal{D}_\delta}\right)\left(\mathcal{D}_{\delta,i}-\overline{\mathcal{D}_\delta}\right)^t\,,
\ee
where $m$ is the number of mock samples. The overbar denotes the average mock sample. We then calculate the fiducial $\chi^2$--distribution,
\be
\chi^2_{\delta}=\left(\mathcal{D}_{\delta} - \mathcal{D}_{\text{base}}\right) S^{-1}_\delta \left(\mathcal{D}_{\delta} - \mathcal{D}_{\text{base}}\right)^t \,,
\ee
and, for each $\delta$, we find
\be
\chi^2_{\rm test, \delta} = \left(\mathcal{D}_{\text{test}}-\mathcal{D}_{\text{base}}\right)\ S^{-1}_\delta \left(\mathcal{D}_{\text{test}}-\mathcal{D}_{\text{base}}\right)^t\,.
\ee
We iterate $\delta^\mathcal{D}$ to find a value for which $\chi^2_{\rm test, \delta}$ is the maximum of $\chi^2_{\delta}$, for $\delta \leq 20\%$. Finally, we find $\sigma_\delta = (\delta_+-\delta_-)/2$, where $\delta_+$ is the value that makes $\chi^2_{\rm test}$ fall at the right-hand border of the 68\% probability interval of the $\chi^2$--distribution, and similarly for~$\delta_-$. The steps for finding $\delta^\mathcal{C}$ are analogous to what we have described.

We developed a comprehensible code in python\footnote{\href{https://github.com/t-ferreira/Covariance_comparison}{github.com/t-ferreira/Covariance\_comparison}} that uses this procedure to test $\mathcal{D}$ and $\mathcal{C}$. We found convergence for $m > 5000$. The value of $\delta^{\mathcal{D}}_{\rm test} \pm \sigma_\delta$ estimates the distance between the two compressed matrices, $C_{\text{base}}$ and $C_{\text{test}}$, as far as the uncertainties or correlations in the parameters are concerned.

As Eqs.~\eqref{diago} and~\eqref{corre} show, we adopted a very agnostic point of view: we look for the strength $\delta$ of the Gaussian noise $\mathcal{E}$ that makes one matrix similar to the other.
This zero-mean noise is supposed to arise because of the different choices that are made in the modelling of the covariance matrices.
It is worth stressing that we are not quantifying the bias that may separate the two matrices, which, given the many elements, would be difficult to tackle in a systematic way.

\subsection{Cosmic shear covariance matrices}

\Sfig{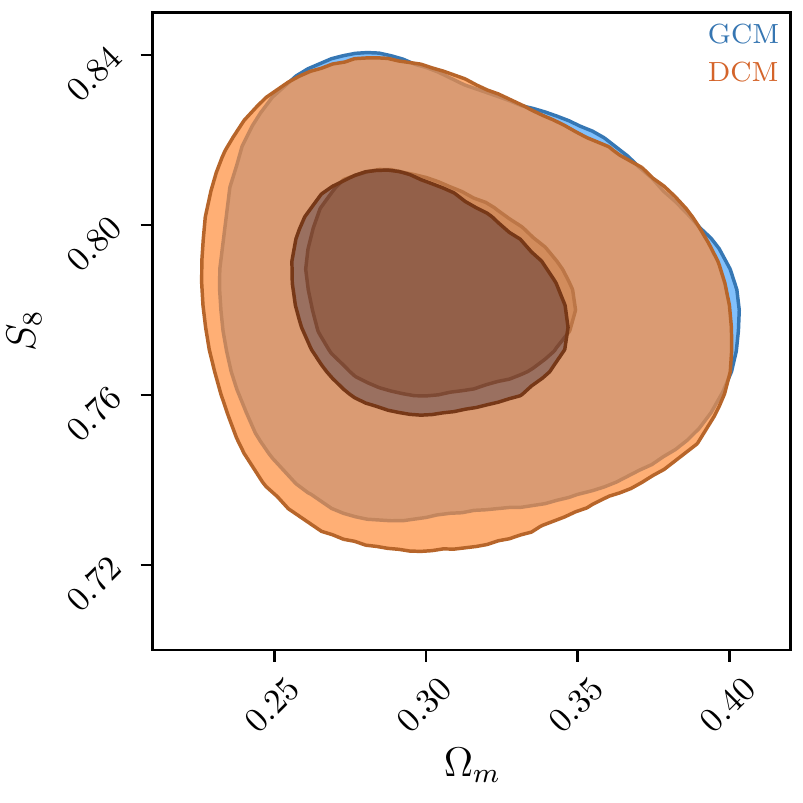}{Constraints on the parameters $\Omega_m$ and $S_8$ for the DCM and GCM produced for cosmic shear, as described in~\rsssec{covariances}. The orange curve is for the DCM, while the blue is for the GCM. In the 16--dimensional parameter space, the volume of the posterior is about $20\%$ smaller for the latter.}
\Sfig{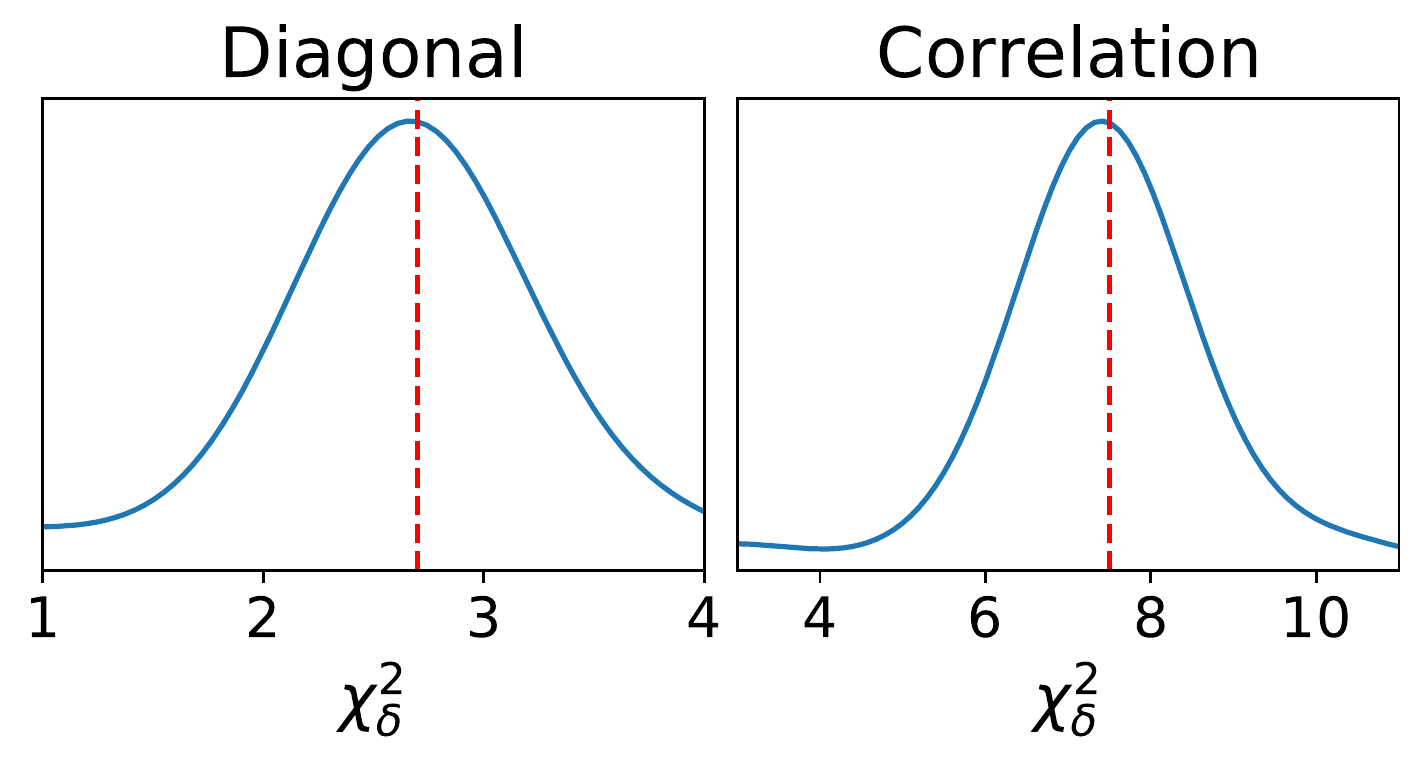}{$\chi^2$--distributions that are used to determine $\delta_{\rm test}$ for the diagonal elements of the covariance matrix (left) and for the elements of the correlation matrix (right).
The method finds the $\delta_{\rm test}$ value such that $\chi^2_{\rm test}$ (red dashed line) falls at the maximum of the distribution.}

We test this method by comparing the DCM and GCM covariance matrices.
We find that the diagonal elements of the compressed GCM differ from the ones of the compressed DCM by
\begin{align} \label{D-meto}
    \delta^{\mathcal{D}}_{\rm test} = 2.6 \pm 0.5\% \,,
\end{align}
and the correlations by
\begin{align} \label{C-meto}
    \delta^{\mathcal{C}}_{\rm test} = 7.5 \pm 0.6\% \,.
\end{align}
We illustrate these findings in \rf{Errors}, which shows the $\chi^2$--distributions used by our method: the values of $\delta_{\rm test}$ are such that the $\chi^2_{\rm test}$ values fall at the maximum of the distributions.

In order to validate these results, we carry out a full Bayesian analysis as shown in~\rf{wmS8-Y1}, where one can see that the two matrices give similar constraints.
We then estimate from the chains the two $16\times16$ covariance matrices on the parameters, that is, the second moments of the posterior distribution.
These moments give the Gaussian errors that are relative to the full (possibly non-Gaussian) posterior.
In order to compare with the results of Eqs.~(\ref{D-meto}-\ref{C-meto}), which are based on the Fisher matrices on the parameters, we then extract the diagonal elements $\mathcal{D}_{\rm DCM}$ and $\mathcal{D}_{\rm GCM}$ and the correlation vectors $\mathcal{C}_{\rm DCM}$ and $\mathcal{C}_{\rm GCM}$ from the inverse of the covariance matrices of the DCM and GCM chains.
Note that the latter matrices are different from the Fisher matrix, which is only sensitive to the maximum of the likelihood.
Next, we define:
\begin{align}
\mathcal{E}_{\mathcal{D}} &= \sqrt{ \frac{\mathcal{D}_{\rm GCM}}{\mathcal{D}_{\rm DCM}}}-1 \,, \\
\mathcal{E}_{\mathcal{C}} &= \left (1- \mathcal{C}_{\rm GCM}^2 \right) \left ( \tanh^{-1} \mathcal{C}_{\rm GCM} - \tanh^{-1} \mathcal{C}_{\rm DCM} \right ) \,,
\end{align}
and compute the standard deviations of $\mathcal{E}_{\mathcal{D}}$ and $\mathcal{E}_{\mathcal{C}}$ to see if their values agree with the results of Eqs.~(\ref{D-meto}-\ref{C-meto}). 
We find:
\begin{align}
\sigma_{\mathcal{E}_{\mathcal{D}}} & = 1.7 \% \,, \label{D-val}\\
\sigma_{\mathcal{E}_{\mathcal{C}}} & = 1.5 \% \,. \label{C-val}
\end{align}
We find a $2\sigma$ agreement for the diagonal elements and that our method overestimates the difference in the correlations. The latter is somewhat expected as correlations are more difficult to quantify when close to zero and for variables that are strongly degenerated because of non-Gaussianities in the posterior.
Regarding the latter effect we stress again that the results of Eqs.~(\ref{D-meto}-\ref{C-meto}) are based on the Fisher matrices on the parameters, while the ones of Eqs.~(\ref{D-val}-\ref{C-val}) are based on the full posterior which may be significantly non-Gaussian in, especially, the nuisance parameters.

Finally, our algorithm took roughly 0.5~CPUh (2020 laptop) to generate the desired output, while a full cosmological analysis takes roughly 100 times more.


\section{Conclusion}
\label{sec:conclusion}

The complexity and the considerable size of covariance matrices of large datasets make them increasingly difficult to analyse. Because of the vast range of values, with their own elements often differing by several orders of magnitude, it is customary to identify the largest elements and the diagonal ones. Should they be similar, then it is likely that the parameter constraints are also compatible; we show here that this is a perilous assumption. Using an invertible transformation, we showed that it is possible to generate very distinct covariance matrices capable of generating the same cosmology, given the same data vector.

On the other hand, we follow the groundwork established by~\cite{Ferreira2020}, which showed MOPED is capable of reducing the size of the covariance matrix while retaining the necessary information for parameter constraints for cosmic shear statistics. We thus use the compressed matrix formalism to develop a fast and reliable method for the comparison that uses compressed covariance matrices. Together with this work we release a comprehensible implementation in python of this method, which is available at \href{https://github.com/t-ferreira/Covariance_comparison}{github.com/t-ferreira/Covariance\_comparison}.

A direct one-to-one element comparison, as is usually done, by considering the perturbed covariance matrices generated in Section~\ref{sec:rotation} would have led one to erroneously expect notable differences in their constraining power as compared to those obtained with the original covariance matrix. This incorrect assumption is avoided if considering MOPED-compressed covariance matrices, since they are identical. This approach thus ensures an equitable comparison since we limit ourselves to the most relevant elements. Using our method, we show that the DESY1 covariance matrix and the Gaussian covariance matrix, which produce compatible cosmological constraints, feature a difference of $7.5\%$ for the elements of their correlation matrix and $2.6\%$ for their diagonal elements of the covariance matrix.
The latter figure is in agreement with the results from the full Bayesian analysis which gives a 1.7\% difference.
Regarding the correlations, the full analysis gives a lower value, $1.5\%$, possibly because of the difficulty in estimating the correlation for degenerate cases.
While our approximate method cannot replace a full Bayesian analysis, it is a powerful and reliable tool for the development and validation of codes that estimate covariance matrices. Our method takes roughly 100 times less CPUh than a full cosmological analysis.

Finally, it is relevant to point out that the comparison presented here is done between analytical covariance matrices, and that drawing $\delta$ from a Gaussian distribution proved suitable for reproducing the differences between their parameter constraints. In the case of covariance matrices obtained using different methods, such as comparing an analytical and a simulated one, the Wishart distribution is a likely better candidate for generating $\delta$~\citep{Taylor2013}.

\section*{Acknowledgements}

It is a pleasure to thank Scott Dodelson for useful comments and discussions. TF also thanks Tilman Troester for helpful comments on how to use KCAP.
TF thanks CAPES and FAPES for financial support.
VM thanks CNPq and FAPES for partial financial support.

\section*{Data Availability}

The algorithm used as well as the data underlying this article can be found in \href{https://github.com/t-ferreira/Covariance_comparison}{github.com/t-ferreira/Covariance\_comparison}, where we have included a notebook example to reproduce our results.


\bibliographystyle{mnras}
\bibliography{main}



\appendix

\section{Cosmic shear statistics}
\label{ap:shear}

The mapping of the light distribution of the source to image coordinates is done via the magnification tensor, which is the inverse of the Jacobian matrix,
\be
    \mathbf{A} = \left(\begin{matrix}
    1 - \kappa - \gamma_1 & - \gamma_2 \\[1.1ex]
    - \gamma_2 & 1 - \kappa + \gamma_1\\[1.1ex]
    \end{matrix}\right)\ ,
\ee
with $\kappa$ being the isotropic decrease or increase of the observed size of the source image and the anisotropic deformation $\gamma = \gamma_1 + i \gamma_2$ being the shear. In Fourier space, these two quantities are related via
\be\eql{gamma_kappa_relation}
    \Tilde{\gamma} (\bm{\ell}) = \frac{(\ell_1 + i\ell_2)^2}{\ell^2} \Tilde{\kappa}(\bm{\ell}) = \text{e}^{2i\beta}\Tilde{\kappa}(\bm{\ell})\ ,
\ee
where $\beta$ is the polar angle of the wave vector $\bm{\ell} = (\ell_1,\ell_2)$. 

The convergence can also be interpreted according to the projected matter density, with its power spectrum being associated with that of the matter density contrast, $P_\delta$~\citep{Schneider2002}. Using several approximations, such as the Limber projection, which uses only modes that lie in the plane of the sky, the small-angle approximation, and the flat-sky limit, where we replace spherical harmonics by Fourier transforms, we can write,
\be
    P_{\kappa}(\ell) = \frac{9}{4}\Omega_m^2 \left(\frac{H_0}{c}\right)^4 \!\!\!\! \int^{\chi_{\textrm{lim}}}_{0}\!\!\!\!\! d\chi \frac{g^2(\chi)}{a^2(\chi)} P_{\delta}\!\! \left(\! k \!=\! \frac{\ell}{f_{K}(\chi)}, \chi \! \right)\ .
\ee
The integral goes to the limiting comoving distance of the galaxy sample, $\chi_{\textrm{lim}}$, and we identify $H_0$ as the value of the Hubble constant today, $c$ as the speed of light, $\chi$ as the comoving coordinate, $a$ as the scale factor, $f_K$ as the comoving angular distance, and the reduced shear, $g$, given by,
\be
    g = \frac{\gamma}{1 - \kappa}\ .
\ee

The real-space shear two-point correlation function is the main cosmic shear observable because it can be readily obtained by averaging over the multiplied ellipticities of galaxy pairs. We can decompose the shear into its tangential component, $\gamma_t$, and the cross-component, $\gamma_{\times}$, which are defined as
\be
    \gamma_t = - \text{Re} \left(\gamma \text{e}^{-2\text{i}\phi}\right) \  , \quad \gamma_{\times} = - \text{Im} \left(\gamma \text{e}^{-2\text{i}\phi}\right)\ ,
\ee
where $\phi$ is the polar angle of the separation vector $\bm{\theta}$. The shear correlation functions are then,
\bea
    \xi_{+} (\theta) &= \langle \gamma \gamma^* \rangle (\theta) = \langle \gamma_t \gamma_t \rangle (\theta) + \langle \gamma_{\times} \gamma_{\times} \rangle (\theta) \ , \\
    \xi_{-} (\theta) &= \text{Re} \left(\langle \gamma \gamma^* \rangle (\theta)\text{e}^{-4\text{i}\phi}\right) = \langle \gamma_t \gamma_t \rangle (\theta) - \langle \gamma_{\times} \gamma_{\times} \rangle (\theta) \ .
\eea
Its power spectrum is given by the Fourier transform of the correlation function,
\bea
    \langle\Tilde{\gamma} (\bm{\ell})\Tilde{\gamma}^* (\bm{\ell}')\rangle & = \left( 2\pi\right)^2 \delta_D\left(\bm{\ell} - \bm{\ell}' \right)P_\kappa(\ell)\ , \\
    \langle\Tilde{\gamma} (\bm{\ell})\Tilde{\gamma} (\bm{\ell}')\rangle & = \left( 2\pi\right)^2 \delta_D\left(\bm{\ell} - \bm{\ell}' \right)\text{e}^{4\text{i}\phi}P_\kappa(\ell)\ ,
\eea
with $\delta_D$ being the Dirac function and using \ec{gamma_kappa_relation} to write $P_\kappa(\ell) = P_\gamma(\ell)$. Finally, if we choose $\bm{\theta} = (\theta, 0)$, we have,
\bea
    \xi_{+} (\theta) &= \langle \gamma (\mathbf{0}) \gamma^* (\bm{\theta}) \rangle \\
    &= \int \frac{d^2\ell}{\left( 2\pi\right)^2}\int \frac{d^2\ell'}{\left( 2\pi\right)^2} \text{e}^{\text{i}\bm{\ell}'\cdot\bm{\theta}} \langle\Tilde{\gamma} (\bm{\ell})\Tilde{\gamma}^* (\bm{\ell}')\rangle \\
    &= \int \frac{d^2\ell'}{\left( 2\pi\right)^2} \text{e}^{\text{i}\bm{\ell}'\cdot\bm{\theta}} \int d^2\ell\ \delta_D\left(\bm{\ell} - \bm{\ell}' \right)P_\kappa(\ell) \\
    &= \int \frac{d^2\ell}{\left( 2\pi\right)^2} \text{e}^{\text{i}\bm{\ell}\cdot\bm{\theta}} P_\kappa(\ell)\ ,
\eea
where we make the substitution $\bm{\ell}' \rightarrow \bm{\ell}$. If we expand $\int d^2\ell = \int \ell d\ell \int d\varphi$,
\be
    = \int_0^\infty \frac{\ell d\ell}{\left( 2\pi\right)^2} P_\kappa(\ell) \int_0^{2\pi} \text{e}^{\text{i}\ell \theta \text{cos}\varphi} d\varphi\ .
\ee
Substituting
\be
    \int_0^{2\pi} \text{e}^{\text{i}\ell \theta \text{cos}\varphi} d\varphi = 2\pi J_0(\ell \theta)\ ,
\ee
we arrive at the well-known result,
\be
    \xi_{+} (\theta) = \int_0^\infty \frac{\ell d\ell}{2\pi} J_0(\ell \theta) P_\kappa(\ell)\ .
\ee
Similarly, we obtain
\be
    \xi_{-} (\theta) = \int_0^\infty \frac{\ell d\ell}{2\pi} J_4(\ell \theta) P_\kappa(\ell)\ .
\ee


\bsp	
\label{lastpage}
\end{document}